\documentclass[12pt]{article}
\usepackage[latin1]{inputenc}
\usepackage[dvips]{epsfig}
\usepackage{latexsym}
\usepackage{dsfont}
\usepackage{a4}

\begin{document}

\title{Multiplication in Cyclotomic Rings and its Application to Finite Fields}
\author{Francisco Arg\"{u}ello \\
Dept. Electr\'{o}nica y Computaci\'{o}n \\
Universidad de Santiago de Compostela. \\
15782 Santiago de Compostela. Spain. \\
francisco.arguello@usc.es  }

\begin{titlepage} 
\maketitle 

\vspace{4cm}
\begin{tabbing} 
{\bf Mailing Address:} \= Francisco Arguello \\ 
\> Dept. Electronica y Computacion \\ 
\> Universidad de Santiago de Compostela \\ 
\> 15782 Santiago de Compostela \\ 
\> Spain \\  
\> \\ 
\> {\bf PHONE:} \= +34 981 594488 ext. 13556 \\ 
\> {\bf FAX:}   \> +34 981 528012  \\ 
\> {\bf e-mail:} \> francisco.arguello@usc.es \\ 
\end{tabbing} 
\end{titlepage} 

\newpage

\begin{center}
{\Large \textbf{Multiplication in Cyclotomic Rings and its Application to Finite Fields}}

\vspace{0.5cm}
Francisco Arg\"{u}ello
\end{center}

\begin{abstract}
A representation of finite fields that has proved useful when implementing finite field arithmetic in hardware is based on an isomorphism between subrings and fields. In this paper, we present an unified formulation for multiplication in cyclotomic rings and cyclotomic fields in that most arithmetic operations are done on vectors. From this formulation we can generate optimized algorithms for multiplication. For example, one of the proposed algorithms requires approximately half the number of coordinate-level multiplications at the expense of extra coordinate-level additions. Our method is then applied to the finite fields GF($q^m$) to further reduce the number of operations. We then present optimized algorithms for multiplication in finite fields with type-I and type-II optimal normal bases.
\end{abstract}

\vspace{0.5cm}
\textbf{Keywords:} Cyclotomic ring, Finite field, Galois field, Normal basis, Redundant basis, Multiplier.

\section{Introduction}
Recently, there has been a good deal of interest in developing hardware and software methods for implementing the finite field GF($q^m$) arithmetic operations particularly for cryptographic applications \cite{hasan}, \cite{moon}, \cite{leong}. Multiplication in finite fields is a complicated and time-consuming operation that very much depends on how the field elements are represented. A representation of finite fields that has proved useful when implementing finite field arithmetic in hardware is based on an isomorphism between subrings and fields. The main idea is to embed a field in a larger ring, perform multiplication there, and then convert the result back to the field. The ring used is referred to as {\it cyclotomic}, because has an extremely simple basis whose elements form a cyclic group. Because the dimension of the ring is higher than that of the field, this representation is referred to as redundant. Having in mind the design of efficient arithmetic circuits, it is desirable to find the ring of lowest dimension with the property that the finite field is contained in the ring. This way of representation of finite fields has been explored by various authors \cite{itoh}, \cite{drolet}, \cite{geisel}, \cite{gao1}, \cite{gao2}, \cite{silverman}, \cite{wu}.

Drolet \cite{drolet} represents the finite field GF($2^m$) as a subring of the {\it cyclotomic ring} $\mbox{GF}(2)[x]/(x^n+1)$ with the integer $n$ chosen in such a way that $x^n+1\in \mbox{GF}(2)[x]$ contains an irreducible factor of degree $m$. He shows that this ring representation of elements of the finite field satisfies a generalized Massey-Omura condition and the square of an element can be obtained by applying a specific permutation to the bits of the word representing it. In this line, Geiselmann et al. \cite{geisel} characterize the smallest $n$ with $\mbox{GF}(2)[x]/(x^n+1)$ containing an isomorphic copy of GF($2^m$). 


Some redundant bases can be easily introduced by the normal bases generated with the help of a Gauss period \cite{mullin}, \cite{menezes}. Gao et al. \cite{gao1}, \cite{gao2} use Gauss periods for embedding the elements of the finite field in a {\it cyclotomic field} and, by doing so, they can find the relation/conversion between the redundant basis and the normal basis. This conversion can be done in hardware with almost no cost. There are two types of normal basis generated by Gauss periods with minimal complexity, usually called optimal normal bases (ONBs) of type-I and type-II, respectively. When there exists an ONB, very simple and highly regular multiplier architecture can be obtained using the redundant representation. Recently, Wu et al. \cite{wu} have made this idea more explicit and present architectures that are suitable for hardware implementation. The basic idea is to embed the finite field GF($2^m$) in the smallest splitting field of $x^n+1$ over GF(2) and do the arithmetic in this cyclotomic field. 

In this paper, we first present an unified formulation for multiplication in cyclotomic rings and cyclotomic fields in that most arithmetic operations are done on vectors. The method is quite generic in the sense that it is not restricted to any special type of ground field.  Our algorithms are then applied to the finite fields GF($q^m$) with $q$ prime to further reduce the number of operations. The organization of the rest of this paper is as follows: In the next section, we briefly review the cyclotomic rings and fields. In Section 3, we derive a formulation for multiplication in generic cyclotomic rings/fields. We also give the computational complexity of the algorithms in terms of the coordinate-level operations needed. In section 4, we apply the method to the finite fields and then adapt it to two special classes of bases, namely, the type-I and type-II ONBs. Finally, we make a few concluding remarks in Section 5.

\section{Cyclotomic Rings and Fields}

\subsection{Cyclotomic rings}
Let $F$ be a field. The set of polynomials with coefficients in the field, $F[x]$, with the usual operations of addition and multiplication of polynomials forms a ring. We can also consider the ring of the polynomials modulo a polynomial $p(x)$. If we let $\beta$ be the residue class of $x$, then the elements of $F[x]/p(x)$ can be represented in the form 

\begin{equation}
A=a_0+a_1\beta+a_2\beta^2+\cdots+a_{n-1}\beta^{n-1},~~a_i\in F, \label{repre}
\end{equation}
where $n$ is the degree of $p(x)$. That is, the elements $1,\beta,\beta^2,\ldots,\beta^{n-1}$ form a true basis for $F[x]/p(x)$.

If the arithmetic is done modulo the polynomial $x^n-1$ then one obtains the $n$th {\it cyclotomic ring} $F[x]/(x^n-1)$. Since a cyclotomic ring satisfies the expression $\beta^n= 1$, the elements $1,\beta,\beta^2,\ldots,\beta^{n-1}$ form a cyclic group of order $n$ with the following multiplication table:

\begin{equation}
\beta\cdot\beta^i =\left\{ \begin{array}{ll} \beta^{i+1} & \mbox{ if } 0\leq i< n-1 \\
1  & \mbox{ if } i = n-1 . \end{array}\right . \label{basica}
\end{equation}

As mentioned in the introduction, the key idea of the representation of GF($q^m$) considered in \cite{drolet}, \cite{geisel} is to represent the field GF($q^m$) as a subring of $\mbox{GF}(q)[x]/(x^n-1)$ with $n\geq m$ and do the arithmetic operations over the ring.

\begin{description}
\item{\bf Example 1.} 
With the usual addition and multiplication in $\mbox{GF}(2)[x](x^3+1)$, the residue classes 0, $\beta+1$, $\beta^2+1$ and $\beta^2+\beta$ form a subring of $\mbox{GF}(2)[x]/(x^3+1)$ that is isomorphic to GF($2^2$). The residue class $\beta^2+\beta$ serves as a multiplicative identity in the subring.
\end{description}

\subsection{Cyclotomic fields}
On the other hand, the $n$th {\it cyclotomic field} \cite{wu} over the field $F$, denoted $F^{(n)}$, is defined to be the splitting field of $x^n-1$ over $F$. Let $\beta$ be a primitive $n$th root of unity in some extension of $F$. Then, the elements $1,\beta,\beta^2,\ldots,\beta^{n-1}$ form a cyclic group of order $n$ with the multiplication table (\ref{basica}). $F^{(n)}$ is obtained by adjoining the elements generated by $\beta$ to $F$. We may consider the basis $[1,\beta,\beta^2,\ldots,\beta^{n-1}]$ and write elements of $F^{(n)}$ in the form (\ref{repre}).

Since a cyclotomic field satisfies the equation

\begin{equation}
1+\beta+\beta^2+\cdots+\beta^{n-1}=0, \label{betas1}
\end{equation}
the representation is not unique, that is, each $n$-tuple $(a_0,a_1,\ldots,a_{n-1}),~a_i\in F$, gives an element of $F^{(n)}$, but different tuples may give the same element. For example, since (\ref{betas1})
the two $m$-tuples $(a_{0},a_{1},\ldots,a_{n-1})$ and $(a_{0}+k,a_{1}+k,\ldots,a_{n-1}+k),~k\in F$ both represent the same element. 

\begin{description}
\item{\bf Example 2.} 
If $Q$ is the field of rational numbers and $n=3$, then a cyclotomic field $Q^{(3)}$ can be obtained by adjoining a primitive cubic root of unity, $\beta$, say $\beta=(-1+i\sqrt{3})/2$, to the rational numbers $Q$, and the elements of $Q^{(3)}$ can be written as $A=a_0+a_1\beta+a_2\beta^2$. Note that equation (\ref{betas1}) is satisfied and so such representation is redundant since we can also write $A= b_0+b_1i\sqrt{3}$.
\end{description}

As mentioned in the introduction, the basic idea in \cite{wu} is to embed the finite field GF($q^m$) in the smallest splitting field of $x^n-1$ over GF($q$) and do the arithmetic in this cyclotomic field. Some examples are the redundant bases which can be generated with the help of the Gauss periods \cite{menezes}. If there exists a normal basis $[\gamma^{q^0},\gamma^{q^1},\ldots,\gamma^{q^{m-1}}]$ generated by a Gauss periods of type $(m,k)$, then this normal basis can be expressed in function of the redundant basis $[\beta^0,\beta^1,\ldots,\beta^{n-1}]$ as

\begin{equation}
[\gamma^{q^0},\gamma^{q^1},\ldots,\gamma^{q^{m-1}}]=\left[ \sum_{i=0}^{k-1}\beta^{q^0\alpha^i}, \sum_{i=0}^{k-1}\beta^{q^1\alpha^i}, \ldots ,\sum_{i=0}^{k-1}\beta^{q^{m-1}\alpha^i}\right]. \label{redundante}
\end{equation}
where $n=mk+1$, $\alpha$ is an element of orden $k$ of $\mathds{Z}_n^\times$ and $\beta$ satisfies, by construction, equations (\ref{basica}) and (\ref{betas1}).

\subsection{Multiplication}
Let any two elements $A,B$ be represented in the form (\ref{repre}), i.e., $A=\sum_{i=0}^{n-1} a_i\beta^i$ and $B=\sum_{i=0}^{n-1} b_i\beta^i$. Since $\beta^n=1$ (that is satisfied in both, cyclotomic rings and cyclotomic fields), the product $C=AB$ can be written as,

\begin{equation}
C=\sum_{i=0}^{n-1}\sum_{j=0}^{n-1}a_i b_j \beta^{i+j} =\sum_{j=0}^{n-1}\left(\sum_{i=0}^{n-1}a_i b_{j-i} \right)\beta^j , \label{pcic2}
\end{equation}
where the subscript $j-i$ must be read modulo $n$ (i.e., $a_{n+k}\rightarrow a_k$ and $a_{-k}\rightarrow a_{n-k}$). Then, the coordinates of $C$ can be calculated by

\begin{equation}
c_j=\sum_{i=0}^{n-1}a_i b_{j-i},~0\leq j < n. \label{pcic}
\end{equation}

The resulting algorithm is suitable for a bit-level hardware implementation \cite{wu}.

\section{Algorithm for Multiplication}
In this section, we will introduce a vector-level algorithm which essentially eliminates the bit-wide inner products needed by a direct implementation of equation (\ref{pcic}). We start from the equation (\ref{pcic2}) and, using a similar technique to that of \cite{rey3}, write a separate sum with the terms which have equal coordinate indexes, 

\begin{eqnarray}
C&=& \sum_{i=0}^{n-1}\sum_{j=0}^{n-1}a_i b_j \beta^{i+j} 
=\sum_{i=0}^{n-1}a_i b_i \beta^{2i}
+\sum_{i=0}^{n-1}\sum_{j=0,j\neq i}^{n-1} a_i b_j\beta^{i+j} \\
&=& \sum_{i=0}^{n-1}a_i b_i \beta^{2i}
+\sum_{i=0}^{n-1}\sum_{k=1}^{n-1} a_i b_{i+k}\beta^{2i+k}.
\end{eqnarray}

In the last expression we have used that $\beta^{2i+k}=\beta^{2i+k-n}$ for $2i+k\geq n$ and the subscripts must be read modulo $n$. Denoting $v=\lfloor(n-1)/2\rfloor$ and since the multiplication matrix is symmetric, we can write

\begin{equation}
C= \sum_{i=0}^{n-1}a_i b_i \beta^{2i}+\sum_{i=0}^{n-1}\sum_{j=1}^{v}(a_ib_{i+j}+a_{i+j}b_i)\beta^{2i+j}+V, \label{finalring0}
\end{equation}
where

\begin{equation}
V=\left\{ \begin{array}{ll}
\displaystyle\sum_{i=0}^{n-1}a_ib_{i+n/2}\beta^{2i+n/2}=\sum_{i=0}^v (a_i b_{i+n/2}+ a_{i+n/2}b_i) \beta^{2i+n/2} & \mbox{if $n$ even} \\
\displaystyle 0  & \mbox{if $n$ odd}.
\end{array}\right .
\end{equation}

This equation can be rewritten if we add and subtract the term (for $n$ odd),

\begin{equation}
\sum_{i=0}^{n-1} \sum_{j=0}^{v}a_i b_i\beta^{2i+j}+\sum_{i=0}^{n-1} \sum_{j=0}^{v} a_{i+j}b_{i+j}\beta^{2i+j}.
\end{equation}

The last sum can be re-indexed, and then one can verify that

\begin{equation}
C= \left(1-W\right)\sum_{i=0}^{n-1}a_i b_i \beta^{2i} +\sum_{i=0}^{n-1}\sum_{j=1}^{v}(a_i+a_{i+j})(b_i+b_{i+j})\beta^{2i+j}+Z , \label{desvio}
\end{equation}
where

\begin{equation}
W=\left\{ \begin{array}{ll}
\displaystyle \sum_{j=1}^{v}\left(\beta^j+\beta^{-j}\right)+\beta^{n/2}& \mbox{if $n$ even} \\
\displaystyle \sum_{j=1}^{v}\left(\beta^j+\beta^{-j}\right) & \mbox{if $n$ odd},
\end{array}\right .
\end{equation}
and

\begin{equation}
Z=\left\{ \begin{array}{ll} 
\displaystyle\sum_{i=0}^v (a_i+a_{i+n/2})(b_i+b_{i+n/2}) \beta^{2i+n/2}& \mbox{if $n$ even} \\
\displaystyle 0 & \mbox{if $n$ odd} .
\end{array}\right . 
\end{equation}

We can also write $W=\sum_{j=1}^{n-1}\beta^{j}$. Let $A=\sum_{i=0}^{n-1} a_i \beta^i$ be any ring element. Then one can verify that

\begin{equation}
A (1-W)= \sum_{i=0}^{n-1} (2a_i-p)\beta^i,~~\mbox{with}~p=\sum_{j=0}^{n-1}a_j.
\end{equation}   

Applying this expression to (\ref{desvio}) and by using a bit of algebra, we can obtain, 

\begin{equation}
C= x \sum_{i=0}^{n-1}\beta^i + 2\sum_{i=0}^{n-1}a_i b_i \beta^{2i} +\sum_{i=0}^{n-1}\sum_{j=1}^{v}(a_i+a_{i+j})(b_i+b_{i+j})\beta^{2i+j}+Z , \label{finalring1}
\end{equation}
with $x=-\sum_{j=0}^{n-1} a_j b_j$. This equation applies to both cyclotomic fields and cyclotomic rings and to any ground field $F$.

In the case of cyclotomic fields, we can apply the supplementary relation (\ref{betas1}). So, in this case, we obtain the equation

\begin{equation}
C= 2 \sum_{i=0}^{n-1} a_i b_i \beta^{2i}+\sum_{i=0}^{n-1}\sum_{j=1}^{v}(a_i+a_{i+j})(b_i+b_{i+j})\beta^{2i+j} +Z . \label{finalfield1}
\end{equation}

Equations (\ref{finalring0}), (\ref{finalring1}) and (\ref{finalfield1}) are the final results. In the next section, we will see how to obtain a vector-level algorithm from these equations.

Table 1 compares the number of coordinate-level operations of the obtained equations with that of a direct implementation of equation (\ref{pcic}) (for example, figures 1.a and 1.b \cite{drolet},\cite{geisel},\cite{wu}). Equation (\ref{finalring0}) requires the same number of multiplications and additions as the direct implementation. On the other hand, equations (\ref{finalring1}) and (\ref{finalfield1}) require approximately half the number of coordinate-level multiplications. Although this is achieved at the expense of extra coordinate-level additions, the total number of operations is only slightly higher than that of the direct implementation. Hence, these equations are advantageous for ground fields in which multiplication is more costly than addition. In the particular case where the ground field is GF(2), the number of operations is slightly lower because of $2a_ib_i=0$. In this case, equation (\ref{finalfield1}) requires the lowest number of operations.

\begin{table}
\begin{center}
\begin{tabular}{|l|c|c|c|c|} \hline
Multiplier & \#Mult & \#Doub & \#Add & Total \\ \hline
Eqs. (\ref{finalring0}), (\ref{finalring3}), rings and fields & $n^2$ & 0 & $(n-1)n$ & $2n^2-n$ \\
Eqs. (\ref{finalring1}), (\ref{finalring2}), rings (general)  & $(n+1)n/2$  & $n$ & $(3n+1)n/2-1$  & $2n^2+2n-1$ \\ 
Eqs. (\ref{finalring1}), (\ref{finalring2}), rings (GF(2)) & $(n+1)n/2$  &   0 & $(3n-1)n/2-1$  & $2n^2-1$ \\ 
Eqs. (\ref{finalfield1}), (\ref{finalfield2}), fields (general) & $(n+1)n/2$  & $n$ & $3(n-1)n/2$    & $2n^2$   \\
Eqs. (\ref{finalfield1}), (\ref{finalfield2}), fields (GF(2))   & $(n-1)n/2$  &   0 & $(3n-5)n/2$    & $2n^2-3n$ \\ 
Direct \cite{drolet},\cite{geisel}   & $n^2$       &   0 & $(n-1)n$       & $2n^2-n$\\ \hline
\end{tabular}
\caption{Comparison of cyclotomic ring/field multipliers.}
\end{center}
\end{table}

\section{Application to finite fields}
Finally, we restrict ourselves to the particular case of $n$ odd (for example, for finite fields GF($q$) with $q$ prime or power of a prime). In this case and for both cyclotomic rings and cyclotomic fields, equation (\ref{finalring0}) can be written as

\begin{equation}
C= \sum_{i=0}^{n-1} \left\{a_i b_i+\sum_{j=1}^{v}(a_{i+j}b_{i-j}+b_{i+j} a_{i-j}) \right\}\beta^{2i}, \label{finalring3}
\end{equation}
where we have made the change of variables: $i\rightarrow i+j,~j\rightarrow -2j$. Moreover, for cyclotomic rings, equation (\ref{finalring1}) simplifies to

\begin{equation}
C= \sum_{i=0}^{n-1}\left\{ x + 2 a_i b_i +\sum_{j=1}^{v}(a_{i+j}+a_{i-j})(b_{i+j}+b_{i-j})\right\}\beta^{2i} , \label{finalring2}
\end{equation}
with $x=- \sum_{j=0}^{n-1} a_j b_j$ and $v=(n-1)/2$. Lastly, for cyclotomic fields, equation (\ref{finalfield1}) simplifies to

\begin{equation}
C= \sum_{i=0}^{n-1}\left\{2 a_i b_i+ \sum_{j=1}^{v}(a_{i+j}+a_{i-j})(b_{i+j}+b_{i-j})\right\}\beta^{2i}. \label{finalfield2}
\end{equation}

Where, in the particular case of modulo 2 arithmetic, for example GF($2$), $2 a_ib_i =0$. From these equations we can develop algorithms in which most arithmetic operations are done on vectors instead of bits. We can make the following considerations:

\begin{itemize}
\item The index $i$ runs over all coordinates of the operands, and consequently the operations of multiplication and addition can be done on vectors.
\item The subscript $i+k$ represents a cyclic shift of $k$ positions with respect to the reference index $i$. It is found in the coordinates $a_{i+j}$ and $a_{i-j}$.
\item The square of the basis ($\beta^{2i}$) can simply be performed with a permutation since we can write for $n$ odd,

\begin{eqnarray}
\beta^{2i}|_{i=0}^{n-1} &=& [1,\beta^2,\beta^4,\ldots,\beta^{2(n-2)},\beta^{2(n-1)}]\nonumber\\
~ &=& [1,\beta^2,\beta^4,\ldots,\beta^{n-1}\beta,\beta^3,\ldots,\beta^{n-2}],
\end{eqnarray}
In this last expression, since $\beta^n=1$, we have applied $\beta^{2j}=\beta^{2j-n}$ if $2j\geq n$. Besides, the inverse permutation represents the realization of a square root operation.
\end{itemize}

Table 2 shows the data-flow of the coordinates of the variable $A$ during the computation of equation (\ref{finalfield2}). In this table, we can see the cyclic shifts which have to be done in each cycle and the final permutation for obtaining $C=AB$. If this final permutation is not done, we will obtain $D=\sqrt{AB}$. Thus, from equation (\ref{finalring3}) we have the following algorithm.

\begin{table}
\begin{center}
\begin{tabular}{|r|ccccccc|} \hline
      & $a_1$ & $a_2$ & $a_3$ & $a_4$ & $a_5$ & $a_6$ & $a_0$  \\
$j=1$ & + & + & + & + & + & + & + \\
      & $a_6$ & $a_0$ & $a_1$ & $a_2$ & $a_3$ & $a_4$ & $a_5$ 
\\ \hline
      & $a_2$ & $a_3$ & $a_4$ & $a_5$ & $a_6$ & $a_0$ & $a_1$  \\
$j=2$ & + & + & + & + & + & + & + \\
      & $a_5$ & $a_6$ & $a_0$ & $a_1$ & $a_2$ & $a_3$ & $a_4$  
\\ \hline
      & $a_3$ & $a_4$ & $a_5$ & $a_6$ & $a_0$ & $a_1$ & $a_2$  \\
$j=3$ & + & + & + & + & + & + & + \\
      & $a_4$ & $a_5$ & $a_6$ & $a_0$ & $a_1$ & $a_2$ & $a_3$  
\\ \hline\hline
      & $\downarrow$ & $\downarrow$ & $\downarrow$ & $\downarrow$ & $\downarrow$ & $\downarrow$ & $\downarrow$ \\ 
$D=\sqrt{AB}\leftarrow$ & $d_0$ & $d_1$ & $d_2$ & $d_3$ & $d_4$ & $d_5$ & $d_6$ \\
$C=AB\leftarrow$        & $c_0$ & $c_2$ & $c_4$ & $c_6$ & $c_1$ & $c_3$ & $c_5$ \\ \hline
\end{tabular}  
\end{center}
\caption{Data-flow of the coordinates of $A$ during the computation of equation (\ref{finalfield2}) for $C=AB$ and $D=\sqrt{AB}$ with $n=7$.}
\end{table}

\vspace{0.5cm}
\noindent\noindent{\bf Algorithm 1.} Multiplication over cyclotomic rings and fields with $n$ odd (equation (\ref{finalring3})).\\ 
{\bf Input:} $A, B$ \\
{\bf Output:} $D=\sqrt{AB},~C=AB$ 
\begin{tabbing}
1. \indent \= Initialize $S_A=A,~ S_B=B, ~v=(n-1)/2$ \\
2.         \> $D= A\odot B$ \\
3.         \> For $j=1$ to $v$ ~~\{ \\
4.         \> \indent \= $S_A<<1,~S_B>>1$ \\
5.         \>         \> $R=(A\odot S_B)\oplus (B\odot S_A)$ \\
6.         \>         \> $D=D\oplus R$ ~~\} \\
7.         \> $C=\mbox{sqrt\_perm}(D)$.
\end{tabbing}

From equations (\ref{finalring2}) and (\ref{finalfield2}) we obtain the following algorithm.

\vspace{0.5cm}
\noindent\noindent{\bf Algorithm 2.} Multiplication over cyclotomic rings (equation (\ref{finalring2})) and cyclotomic fields (equation (\ref{finalfield2})) with $n$ odd. \\ 
{\bf Input:} $A, B$ \\
{\bf Output:} $D=\sqrt{AB},~C=AB$ 
\begin{tabbing}
1. \indent \= Initialize $S_A=A,~ S_B=B,~v=(n-1)/2$ \\
2.         \> $D= A\odot B$ \\
//         \> Lines 3 and 5 apply only to cyclotomic rings \\
3.         \> $x=-\sum_{i=0}^{n-1}d_i$ \\
4.         \> $D=2D$ \\
5.         \> $D=D\oplus (x,\ldots,x)$  \\
6.         \> For $j=1$ to $v$ ~~\{ \\
7.         \> \indent \= $S_A<<1,~S_B>>1$ \\
8.         \>         \> $R=(A\oplus S_A)\odot (B\oplus S_B)$ \\
9.         \>         \> $D=D\oplus R$ ~~\} \\
10.        \> $C=\mbox{sqrt\_perm}(D)$.
\end{tabbing}

In the above algorithms, $\odot$ and $\oplus$ denote coordinate-wise operations, for example, $A\odot B=(a_0b_0,a_1b_1,\ldots,a_{n-1}b_{n-1})$, symbols $<<$ and $>>$ denote cyclic shifts, and $C=\mbox{sqrt\_perm}(D)$ denotes the permutation of coordinates given by $c_{2i{\rm mod}n}=d_i,~0\leq i<n$.

Next, we apply the above algorithms to multiplication in the finite field GF($q^m$). Three cases are considered: the general case (cyclotomic rings), and the two particular cases of finite fields with type-I and type-II ONBs. 

\subsubsection*{A) Cyclotomic rings}
This is the more general case (fields are rings with multiplicative inverses) and so we must use Algorithm 1 or Algorithm 2 in full. 

\begin{description}
\item{\bf Example 3.} GF($2^2$) is isomorphic to a subring of $\mbox{GF}[x]/(x^3+1)$ and so a finite field element can be written in the redundant representation as $A=a_0+a_1\beta+a_2\beta^2$. An isomorphism is given by the embedding $0\rightarrow 0$, $1 \rightarrow \beta+\beta^2$, $\alpha \rightarrow 1+\beta$ and $1+\alpha \rightarrow 1+\beta^2$, where the former is an element of GF($2^2$) in polynomial representation and the latter a ring element. Table 3 shows the operations that must be done in a multiplication when using this redundant representation.
\end{description}

\begin{table}
\begin{center}
\begin{tabular}{|c|c|c|c|} \hline
       & $a_0b_0$              & $a_1b_1$             & $a_2b_2$             \\ 
$j=1$  & $+(a_1 b_2+b_1 a_2)$  & $+(a_2 b_0+b_2 a_0)$ & $+(a_0 b_1+b_0 a_1)$  \\
$C=AB$ & $=c_0$                & $=c_2$               & $=c_1$               \\ \hline
\end{tabular}
\\
(a) \\
\begin{tabular}{|c|c|c|c|} \hline
& \multicolumn{3}{c|} {$x=a_0b_0+a_1b_1+a_2b_2$} \\ \hline
       & $x+$              & $x+$              & $x+$              \\ 
$j=1$  & $(a_1+a_2)$       & $(a_2+a_0)$       & $(a_0+a_1)$       \\
       & $\times(b_1+b_2)$ & $\times(b_2+b_0)$ & $\times(b_0+b_1)$ \\
$C=AB$ & $=c_0$            & $=c_2$            & $=c_1$            \\ \hline
\end{tabular}
\\
(b) 
\end{center}
\caption{Multiplication in a cyclotomic ring (example 3). (a) Algorithm 1. (b) Algorithm 2.}
\end{table}

\subsubsection*{B) ONB-I}
Some cyclotomic fields can be easily introduced by the normal bases generated by the Gauss periods and, by doing so, one can find the relation/conversion between the redundant basis and the normal basis. In these cases, we can used Algorithms 1 and 2 (the latter without lines 3 and 5). 

A type-I ONB can be always generated by a Gauss period of type $(m,1)$. This case is considered in  \cite{gao1}, \cite{gao2}, \cite{silverman}, \cite{wu}, \cite{wang}, \cite{rey2}, \cite{rey3}, \cite{koc}, \cite{rey4}. Here, $n=m+1$, and a basis for GF($q^m$) is $[\beta,\beta^2,\ldots,\beta^{m}]$ (which is a permutation of the normal basis $[\beta^{q^0},\beta^{q^1},\ldots,\beta^{q^{m-1}}]$). The correspondence between finite field elements and cyclotomic field elements is given by

\begin{equation}
\begin{array}{l}
 a_1 \beta +a_2\beta^2+\cdots+a_{m}\beta^{m} ~\longrightarrow~ 0\cdot 1+a_1\beta+a_2\beta^2+\cdots+a_{m}\beta^m, \\
 (a_1-a_0)\beta+(a_2-a_0)\beta^2+\cdots+(a_m-a_0)\beta^m \longleftarrow\\
 ~~~~~~~~~~~~~~~~~~~~~~~\longleftarrow~ a_0\cdot 1+a_1\beta+a_2\beta^2+\cdots+a_{m}\beta^m. \label{map1}
\end{array}
\end{equation}

\begin{description}
\item{\bf Example 4.} The Gauss period $(4,1)$ generates an embedding of GF($2^4$) in the cyclotomic field $\mbox{GF}^{(5)}$. Table 4 shows the operations that must be done in a multiplication using Algorithms 1 and 2.
\end{description}

\begin{table}
\begin{center}
\begin{tabular}{|c|c|c|c|c|c|} \hline
      & $a_0b_0$   & $a_1b_1$   & $a_2b_2$   & $a_3b_3$   & $a_4b_4$ \\
$j=1$ & $+(a_1b_4+b_1a_4)$ & $+(a_2b_0+b_2a_0)$ & $+(a_3b_1+b_3a_1)$ & $+(a_4b_2+b_4a_2)$ & $+(a_0b_3+b_0a_3)$ \\
$j=2$ & $+(a_2b_3+b_2a_3)$ & $+(a_3b_4+b_3a_4)$ & $+(a_4b_0+b_4a_0)$ & $+(a_0b_1+b_0a_1)$ & $+(a_1b_2+b_1a_2)$ \\
$C=AB$& $=c_0$     & $=c_2$     & $=c_4$     & $=c_1$     & $=c_3$ \\ \hline
\end{tabular}
\\
(a) \\
\begin{tabular}{|c|c|c|c|c|c|} \hline
$j=1$ & $(a_1+a_4)$       & $(a_2+a_0)$       & $(a_3+a_1)$       & $(a_4+a_2)$       & $(a_0+a_3)$ \\
      & $\times(b_1+b_4)$ & $\times(b_2+b_0)$ & $\times(b_3+b_1)$ & $\times(b_4+b_2)$ & $\times(b_0+b_3)$ \\ 
$j=2$ & $+(a_2+a_3)$      & $+(a_3+a_4)$      & $+(a_4+a_0)$      & $+(a_0+a_1)$      & $+(a_1+a_2)$      \\
      & $\times(b_2+b_3)$ & $\times(b_3+b_4)$ & $\times(b_4+b_0)$ & $\times(b_0+b_1)$ & $\times(b_1+b_2)$ \\ 
$C=AB$& $=c_0$            & $=c_2$            & $=c_4$            & $=c_1$            & $=c_3$             \\ \hline
\end{tabular}
\\
(b) 
\end{center}
\caption{Multiplication in a type-I ONB (example 4). (a) Algorithm 1. (b) Algorithm 2.}
\end{table}

Equations (\ref{finalring3}) and (\ref{finalfield2}) can be simplified in this case since $a_0=b_0=0$. We can also subtract $c_0$ to $c_1,c_2,...,c_m$ in accordance with the mapping (\ref{map1}). So equation (\ref{finalring3}) simplifies to

\begin{equation}
C'= \sum_{i=1}^m \left\{r+ a_i b_i+ \sum_{\scriptsize\begin{array}{c} j=1, \\j\neq i, j\neq m+1-i\end{array}}^{v} (a_{i+j}b_{i-j}+b_{i+j}a_{i-j})\right\}\beta^{2i}, \label{onb1b}
\end{equation}
where $r=-\sum_{j=1}^{v} a_j b_{m+1-j}+b_j a_{m+1-j}$ and $C'=\sum_{i=1}^m (c_i-c_0)\beta^i$. Also, equation (\ref{finalfield2}) simplifies to

\begin{equation}
C'= \sum_{i=1}^m \left\{ t+ 2a_ib_i+a_{2i} b_{2i} +\sum_{\scriptsize\begin{array}{c} j=1, \\j\neq i, j\neq m+1-i\end{array}}^{v}(a_{i+j}+a_{i-j})(b_{i+j}+b_{i-j})\right\}\beta^{2i}, \label{onb1a}
\end{equation}
where $t=-\sum_{j=1}^{v}(a_j+a_{m+1-j})(b_j+b_{m+1-j})$. 

\subsubsection*{C) ONB-II}
A Gauss period of type $(m,2)$ with $n=2m+1$ generates a type-II ONB. This case considered in \cite{wu}, \cite{wang}, \cite{rey2}, \cite{rey4}, \cite{blake}, \cite{sunar}. From equation (\ref{redundante}), $\gamma^{q^i}=\beta^{q^i}+\beta^{2m+1-{q^i}}$, and so, a mapping between finite field elements and cyclotomic field elements can be written as,

\begin{equation}
\begin{array}{l}
a_1 \beta +a_2\beta^2+\cdots+a_{m}\beta^{m} \longrightarrow \\
~\longrightarrow~ 0\cdot 1+a_1\beta+a_2\beta^2+\cdots+a_{m}\beta^m+a_{m}\beta^{m+1}+a_{m-1}\beta^{m+2}+\cdots+ a_1\beta^{2m}, \\
(a_1-a_0) \beta +(a_2-a_0)\beta^2+\cdots+(a_{m}-a_0)\beta^{m} \longleftarrow\\
~\longleftarrow~ a_0+a_1\beta+a_2\beta^2+\cdots+a_{m}\beta^m+a_{m}\beta^{m+1} +a_{m-1}\beta^{m+2}+\cdots+ a_1\beta^{2m} . \label{map2}
\end{array}
\end{equation}

Again, $[\beta^1,\beta^2,\ldots,\beta^{m}]$ is a permutation of the coefficients of the normal basis. In the particular case of GF($2^m$), coordinate $a_0$ in (\ref{map2}) is always zero.

\begin{description}
\item{\bf Example 5.} The Gauss period $(3,2)$ generates an isomorphism between GF($2^3$) and $\mbox{GF}^{(7)}$. The data-flow of the multiplication is showed in Table 5. In this table, we show the pairs of coordinates over which arithmetic operations are to be performed.
\end{description}

\begin{table}
\begin{center}
\begin{tabular}{|c|c|c|c|c|c|c|c|} \hline
$j=1$ &(1,1)&(2,0)&(3,1)&(3,2)&(2,3)&(1,3)&(0,2)\\
$j=2$ &(2,2)&(3,1)&(3,0)&(2,1)&(1,2)&(0,3)&(1,3)\\
$j=3$ &(3,3)&(3,2)&(2,1)&(1,0)&(0,1)&(1,2)&(2,3)\\\hline
$C=AB$& $c_0$ & $c_2$ & $c_3$ & $c_1$ & $c_1$ & $c_3$ & $c_2$  \\ \hline
\end{tabular} \\
(a) \\
\begin{tabular}{|c|c|c|c|} \hline
$j=1$ &(2,0)&(3,1)&(3,2)\\
$j=2$ &(3,1)&(3,0)&(2,1)\\
$j=3$ &(3,2)&(2,1)&(1,0)\\\hline
$C=AB$& $c_2$ & $c_3$ & $c_1$  \\ \hline
\end{tabular} \\
(b) 
\end{center}
\caption{Data-flow of multiplication in a type-II ONB (example 5). (a) Original. (b) Simplified.}
\end{table}

Representation given by (\ref{map2}) has some redundancies which can be eliminated. First, coordinates $c_1,c_2,\ldots,c_m$ are obtained twice, but it can be easy avoided. Applying this simplification to equation (\ref{finalring3}), we obtain,

\begin{equation}
C''= \sum_{i=0}^{m} \left\{ a_{s(i)} b_{s(i)}+\sum_{j=1}^{m} (a_{s(i+j)}b_{s(i-j)}+b_{s(i+j)}a_{s(i-j)})\right\}\beta^{s(2i)},  \label{simpli2}
\end{equation}
where the indexes must be read modulo $2m+1$, and

\begin{equation}
s(i)=\left\{\begin{array}{ll} 
i & \mbox{~if~} 0\leq i \leq m \\
2m+1-i & \mbox{~if~} m+1\leq i\leq 2m .
\end{array}\right .
\end{equation}

We can also subtract $c_0$ to $c_1,c_2,...,c_m$ in accordance with the mapping (\ref{map2}). Also, since $a_0=b_0=0$, equation (\ref{simpli2}) can be simplified to

\begin{equation}
C'= \sum_{i=1}^{m} \left\{ 2y+a_{s(i)} b_{s(i)}+\sum_{j=1,j\neq i}^{m} (a_{s(i+j)}b_{s(i-j)}+b_{s(i+j)}a_{s(i-j)})\right\}\beta^{s(2i)},  \label{onb2b}
\end{equation}
where $y=-\sum_{j=1}^m a_j b_j$ and $C'=\sum_{i=1}^m (c_i-c_0) \beta^i$, 

Second, since the data-flow matrix of equation (\ref{onb2b}) is symmetric, it is only necessary to compute the diagonal and the upper triangular submatrices. In a similar way, equation (\ref{finalfield2}) can be written as 

\begin{equation}
C'= \sum_{i=1}^{m} \left\{ 4y+2a_ib_i+a_{s(2i)} b_{s(2i)}+\sum_{j=1,j\neq i}^{m} (a_{s(i+j)}+a_{s(i-j)})(b_{s(i+j)}+b_{s(i-j)}) \right\}\beta^{s(2i)}.  \label{onb2a}
\end{equation}

Equations (\ref{onb1b}), (\ref{onb1a}), (\ref{onb2b}) and (\ref{onb2a}) can be applied to GF($q^m$) with $q$ prime. For comparison, we will consider the particular case of GF($2^m$) and other multipliers as shown in Table 6. This table shows the number of bit operations of these multipliers and the time complexity of multipliers in bit-parallel implementation. The multiplier of \cite{wang} is considered to be the first such work published in the open literature, multipliers of \cite{drolet}, \cite{geisel}, \cite{wu} use redundant representation, and those of \cite{rey2}, \cite{hasan2}, \cite{rey4}, \cite{sunar} are more recent work and have the best results among the known existing ones.  

\begin{table}
\scriptsize
\begin{center}
\begin{tabular}{|l|c|c|c|c|} \hline
Multiplier & \#AND & \#XOR & Total & Critical path\\ \hline
Rings, Alg. 1      & $n^2$  & $n^2-n$ & $2n^2-n$ & $T_A+\lceil\log_2 n\rceil T_X$\\
Rings, Alg. 2      & $(n^2+n)/2$ & $(3n^2-n)/2-1$   & $2n^2-1$ & $T_A+(1+\lceil\log_2 n\rceil) T_X$ \\
Rings, redundant \cite{drolet}, \cite{geisel} & $n^2$  & $n^2-n$ & $2n^2-n$ & $T_A+\lceil\log_2 n\rceil T_X$ \\ \hline
ONB-I, Alg. 1      & $m^2+2m+1$  & $m^2+m$        & $2m^2+3m+1$ & $T_A+\lceil\log_2 (m+1)\rceil T_X$\\
ONB-I, Alg. 2      & $(m^2+m)/2$ & $(3m^2+m-2)/2$ & $2m^2+m-1$ & $T_A+\lceil\log_2 (m-1)\rceil T_X$\\ 
ONB-I, Eq. (\ref{onb1b}) & $m^2$ & $m^2-1$ & $2m^2-1$ & $T_A+(1+\lceil\log_2 (m-1)\rceil)T_X$\\
ONB-I, Eq. (\ref{onb1a}) & $(m^2+m)/2$ & $(3m^2-m)/2-1$ & $2m^2-1$ & $T_A+(1+\lceil\log_2 (m-1)\rceil)T_X$\\
ONB-I, redundant \cite{wu}    & $m^2+m$   & $m^2+m$  & $2m^2+2m$ & $T_A+(1+\lceil\log_2 (m-1)\rceil)T_X$\\ 
ONB-I, \cite{wang} & $m^2$ & $2m^2-2m$ & $3m^2-2m$ & $T_A+(1+\lceil\log_2 (m-1)\rceil)T_X$ \\
ONB-I, \cite{hasan2},\cite{rey4} & $m^2$ & $m^2-1$ & $2m^2-1$ & $T_A+(1+\lceil\log_2 (m-1)\rceil)T_X$\\
ONB-I, \cite{rey2} & $(m^2+m)/2$ & $(3m^2-m)/2-1$ & $2m^2-1$ & $T_A+(1+\lceil\log_2 (m-1)\rceil)T_X$ \\\hline
ONB-II, Alg. 1     & $2m^2+m$ & $2m^2$ & $4m^2+m$ & $T_A+\lceil\log_2 (2m+1)\rceil T_X$\\
ONB-II, Alg. 2, Eq. (\ref{simpli2}) & $m^2$ & $3m^2-m$ & $4m^2-m$ & $T_A+(1+\lceil\log_2 m\rceil)T_X$\\
ONB-II, Eq. (\ref{onb2b}) & $m^2$ & $(3m^2-3m)/2$   & $(5m^2-3m)/2$ & $T_A+\lceil\log_2 (2m-1)\rceil T_X$\\ 
ONB-II, Eq. (\ref{onb2a}) & $(m^2+m)/2$ & $2m^2-2m$ & $(5m^2-3m)/2$ & $T_A+\lceil\log_2 (2m-1)\rceil T_X$\\ 
ONB-II, redundant \cite{wu}   & $m^2$ & $2m^2-m$ & $3m^2-m$ & $T_A+(1+\lceil\log_2 m\rceil)T_X$   \\ 
ONB-II, \cite{wang}  & $m^2$ & $2m^2-2m  $ & $3m^2-2m$     & $T_A+(1+\lceil\log_2(m-1)\rceil)T_X$\\
ONB-II, \cite{rey4},\cite{sunar} & $m^2$ & $(3m^2-3m)/2$ & $(5m^2-3m)/2$ & $T_A+(1+\lceil\log_2 m\rceil)T_X$\\ 
ONB-II, \cite{rey2} & $(m^2+m)/2$ & $2m^2-2m$ & $(5m^2-3m)/2$ & $T_A+\lceil\log_2 (2m-1)\rceil T_X$\\ 
\hline
\end{tabular}
\end{center}
\caption{Comparison of GF($2^m$) multipliers.}
\end{table}

For multiplication in finite fields using cyclotomic rings, Algorithm 1 requires the same number of arithmetic operations compared to previously published multipliers, while Algorithm 2 requires approximately half the number of AND operations at the expense of extra XOR operations. For multiplication in finite fields with type-I and type-II ONBs, equations (\ref{onb1b}), (\ref{onb1a}), (\ref{onb2a}) and (\ref{onb2b}) require a total number of operations equal to that of best results known multipliers. 

\section{Conclusions}
In this paper we have presented an unified formulation for multiplication in cyclotomic rings and fields. From this formulation we can generate optimized algorithms for multiplication. The algorithms are quite generic in the sense that they are not restricted to any special type of ground field. Moreover, in our algorithms, most arithmetic operations are done on vectors. One of the proposed algorithms requires approximately half the number of coordinate-level multiplications compared to the conventional algorithm for multiplication in cyclotomic rings/fields. Although this is achieved at the expense of extra coordinate-level additions, the total number of operations is only slightly higher than that of the conventional algorithm. Hence, the proposed algorithm is advantageous for ground fields in which multiplication is more costly than addition. 

Our method has been applied to the finite fields GF($q^m$) to further reduce the number of operations. We then presented optimized algorithms for multiplication in finite fields with ONBs of type-I and type-II. In the particular case of GF($2^m$) and compared to best results known multipliers in GF($2^m$), these proposed ones require the same number of arithmetic operations. 

\section*{Acknowledgments}
This work ~ was ~ supported ~ in part by the Xunta de Galicia ~  under ~ contract PGIDIT03TIC10502PR and by MCYT under contract TIN 2007-67537-C03.

\end{document}